\documentclass[preprint]{aastex}
\usepackage{spr-astr-addons}
\usepackage{url}
\usepackage{rotating}

\RequirePackage{color}

\begin{document}

\title{Application of the MST clustering to the high energy $\gamma$-ray sky. \\
III - New detections of $\gamma$-ray emission from blazars} 

\shorttitle{Minimum Spanning Tree cluster analysis of 5BZCAT sources.}
\shortauthors{R. Campana et al.}

\author{R. Campana}
\affil{INAF--IASF-Bologna, via Piero Gobetti 101, I-40129, Bologna, Italy}
\and
\author{E. Massaro}
\affil{INAF--IAPS, via del Fosso del Cavaliere 100, I-00133, Roma, Italy \\ and \\ In Unam Sapientiam, Roma, Italy}
\and
\author{E. Bernieri}
\affil{INFN--Sezione di Roma Tre, via della Vasca Navale 84, I-00146 Roma, Italy}

\begin{abstract}
We present the results of a photon cluster search in the $\gamma$-ray sky observed by the \emph{Fermi} Large Area Telescope, 
using the new Pass~8 dataset, at energies higher than 10 GeV. 
By means of the Minimum Spanning Tree (MST) algorithm, 
we found 25 clusters associated with catalogued blazars not previously known as $\gamma$-ray emitters. 
The properties of these sources are discussed.
\end{abstract}

  
\section{Introduction}\label{s:introduction}
In two previous papers, 
\citet[][hereafter Paper I]{campana15} and
\citet[][hereafter Paper II]{campana16}, 
we successfully applied a cluster-finding method, 
the Minimum Spanning Tree (hereafter MST) algorithm, to the \emph{Fermi}-LAT sky at energies higher 
than 10 GeV.
The aim was to search new clusters in the arrival directions of $\gamma$-ray photons,
which could be an indication for localized high energy sources. 
These papers illustrated the good performance of MST for finding clusters having a 
small number of photons, but likely related to pointlike sources.

In particular, in Paper I we applied the algorithm to the 6.3 years LAT Pass 7 dataset, 
searching for sources coincident with blazars not previously known as $\gamma$-ray emitters,
because these objects are the majority of identified sources at these high energies
\citep[see the recent review paper by ][]{massaro16}.
The most recent and improved release of the LAT data \citep[Pass 8,][]{atwood13} was used in Paper II 
for an MST search of clusters likely associated with new High Synchrotron Peaked infrared-selected candidate
blazar sources reported in the 1WHSP catalogue \citep{arsioli15}.

In this paper we report the results of a new MST analysis in the same Pass 8 sky 
(above 10 GeV for Galactic latitudes $b > |25\degr|$) which gave a list of 25 clusters located 
very close to blazars reported in the 5th Edition of the Roma-BZCAT (Massaro et al. 2014,
hereafter 5BZCAT) but not in Fermi Collaboration catalogues 
(2FGL, \citealt{nolan12}; 3FGL, \citealt{acero15}; 1FHL, \citealt{1FHL}).
Three of these clusters, however, correspond to sources in the 2FHL catalogue \citep{2FHL}.
As in Paper II we adopted more severe selection criteria with respect to 
those used in the Paper I to obtain a ``safer'' sample.

In Section~\ref{s:mst} the MST algorithm is briefly summarized, 
while in Section~\ref{s:clusters} we discuss its application to the \emph{Fermi}-LAT sky
and the properties of the selected clusters. 
In Section~\ref{s:maxlike} the standard maximum likelihood analysis is applied to the candidate sources, 
in Section~\ref{s:properties} some of these sources are discussed in detail and in Section~\ref{s:discussion} 
we draw our conclusions.

\section{Photon cluster detection by means of the MST algorithm}\label{s:mst}

In this Section a short summary of the MST algorithm principal characteristics is given. 
A more detailed description of the method is presented in Papers I and II, and an extensive
discussion of its statistical properties can be found in \cite{campana08,campana13}. 

The Minimum Spanning Tree algorithm is based on graph theory and searches for clusters in a 
field of $N$ points (\emph{nodes}). 
The MST is the (unique) graph without closed loops, i.e. a \emph{tree}, that connects all the 
nodes with the minimum total weight. 
In our case, the node coordinates are the photon arrival directions on the celestial sphere,
and the edge weights $\lambda_i$ are the angular distances between the locations of photon pairs. 
The total weight is simply the sum of all the edges.

After the computation of the MST, a 2-step \emph{primary selection} extracts a set of subtrees
corresponding to photon clusters.
The first step (\emph{separation}) removes all the edges having a length $\lambda > \Lambda_\mathrm{cut}$, 
the separation value, usually defined in units of the mean edge length 
$\Lambda_m = (\Sigma_i \lambda_i)/N$. 
The resulting set of disconnected sub-trees is further filtered by the second step (\emph{elimination}), 
that removes all the sub-trees having a number of nodes 
$n \leq N_\mathrm{cut}$, 
leaving only the clusters having a size over a properly fixed threshold. 
A first list of clusters is therefore given by remaining set of sub-trees.

A \emph{secondary selection}  is then applied to extract the most robust candidates 
for $\gamma$-ray sources.
A suitable parameter for this selection was found to be the 
\emph{magnitude} of the cluster, defined as
$M_k = n_k g_k  $,
where $n_k$ is the number of nodes in the cluster $k$ and the \emph{clustering parameter} $g_k$ 
is the ratio between $\Lambda_m$ and $\lambda_{m,k}$, the mean length of the $k$-th cluster edges. 
As shown in \cite{campana13}, $\sqrt{M}$ has a linear correlation with other statistical 
significance parameters and that it can be a good estimator of the significance of a MST 
cluster.
Usually \citep[][Papers I, II]{campana13} a lower threshold of $M \sim 20$ is used.

From a weighted mean of the cluster photon's coordinates the \emph{cluster centroid} location can be computed
\citep[see][]{campana13}.
Another quantity that can be evaluated in order to investigate the compatibility of 
the cluster with a $\gamma$-ray source is the radius of the circle centred at the centroid and containing 
the 50\% of photons in the cluster, the \emph{median radius} $R_m$.
For a cluster likely associated with a genuine pointlike 
$\gamma$-ray source, $R_m$ should be smaller than or comparable to the 68\% containment radius of 
instrumental Point Spread Function (PSF).
This radius varies from 0\fdg25 at 3 GeV to 0\fdg12 at 10 GeV in the case of 
front-converting events
\citep{ackermann13b}, using the latest instrumental response 
files\footnote{\url{http://www.slac.stanford.edu/exp/glast/groups/canda/lat_Performance.htm}}. 
We also expect that the angular distance beetween the positions of the cluster centroid 
and the possible optical counterpart are lower than the latter value.

\begin{table*} 
\caption{Coordinates and main properties of MST clusters detected at energies
higher than 10~GeV ($\Lambda_\mathrm{cut} = 0.7\,\Lambda_m$, $N_\mathrm{cut} = 4$) associated with 5BZCAT sources. 
The second and third columns give the J2000 coordinates of MST clusters, angular distances $\Delta \theta$ are computed from these values and the catalogue positions of the 5BZCAT sources. In the last column, \emph{c} indicates a candidate BL Lac in the 5BZCAT, \emph{el} the presence of emission lines in the optical spectrum, \emph{conf} a possible confusion, e.g. the presence of more than one possible counterpart.
The lower section of the table reports the clusters associated with 2FHL sources.
}\label{table1}
\centering
{
\begin{tabular}{crrccrcccl}
\hline

   MST cluster   &     RA   &      DEC  &         5BZCAT source &$\Delta \theta$& $n$& $g$ & $M$ &  $R_m$ & Note \\ 
                   &   J2000  &    J2000  &          &   $'$         &    &   &   &  $\degr$   &  \\
\hline
                   &            &             &          &                          &    &   &   &        &  \\
MST 0216$+$2314 &    34.050 &  23.185   & 5BZB J0216$+$2314 &   5.9 & 13 & 3.466 & 45.058 & 0.117  &  c \\  
MST 0219$-$1725 &    34.767 & $-$17.376 & 5BZB J0219$-$1725 &   1.7 &  9 & 3.162 & 28.460 & 0.113  &  \\  
MST 0314$+$0619 &    48.576 &   6.341   & 5BZB J0314$+$0619 &   1.5 &  7 & 4.837 & 33.856 & 0.033  &  D3PO$^1$ \\ 
MST 0503$-$1115 &    75.889 & $-$11.228 & 5BZB J0503$-$1115 &   1.5 &  6 & 5.024 & 30.146 & 0.043  &  D3PO$^2$ \\ 
MST 0712$+$5719 &   108.171 &  57.337   & 5BZB J0712$+$5719 &   3.0 &  9 & 2.328 & 20.951 & 0.112  &  \\  
MST 0723$+$5841 &   110.893 &  58.677   & 5BZB J0723$+$5841 &   2.7 &  8 & 3.736 & 29.890 & 0.058  &  c, 1FGL$^\dagger$ \\  
MST 0820$+$2353 &   125.237 &  23.889   & 5BZB J0820$+$2353 &   1.4 &  6 & 3.708 & 22.247 & 0.037  &  \\
MST 0828$+$2312 &   126.985 &  23.161   & 5BZB J0828$+$2312 &   2.9 &  8 & 2.854 & 22.836 & 0.081  &  \\
MST 0913$+$8133 &   138.496 &  81.552   & 5BZB J0913$+$8133 &   1.3 &  9 & 2.745 & 24.704 & 0.070  &  \\   
MST 1055$-$0126 &   163.910 &  $-$1.439 & 5BZB J1055$-$0126 &   1.0 & 10 & 6.300 & 63.003 & 0.058  & D3PO$^3$ \\   
MST 1105$+$3946 &   166.469 &  39.747   & 5BZG J1105$+$3946 &   2.1 &  6 & 8.265 & 49.588 & 0.024  &  \\
MST 1134$-$1729 &   173.667 & $-$17.482 & 5BZB J1134$-$1729 &   1.1 &  9 & 2.522 & 22.700 & 0.094  &  \\  
MST 1208$+$3015 &   182.000 &  30.296   & 5BZB J1208$+$3015 &   2.1 &  5 & 4.073 & 20.366 & 0.038  &  \\
MST 1215$+$0732 &   183.767 &   7.516   & 5BZG J1215$+$0732 &   2.0 &  5 & 5.547 & 27.735 & 0.018  &  \\ 
MST 1402$+$1559 &   210.658 &  15.989   & 5BZB J1402$+$1559 &   1.7 &  5 & 4.031 & 20.153 & 0.042  &  el \\
MST 1449$+$2746 &   222.389 &  27.780   & 5BZG J1449$+$2746 &   0.5 &  9 & 2.717 & 24.450 & 0.094  &  conf \\ 
MST 1506$-$0540 &   226.623 &  $-$5.677 & 5BZB J1506$-$0540 &   1.9 &  6 & 5.584 & 33.501 & 0.018  &  \\ 
MST 1515$+$2426 &   229.000 &  24.395   & 5BZG J1515$+$2426 &   2.8 &  7 & 3.488 & 24.416 & 0.062  &  \\ 
MST 1518$+$4045 &   229.637 &  40.737   & 5BZG J1518$+$4045 &   1.4 &  7 & 5.382 & 37.671 & 0.035  &  \\ 
MST 2148$-$0733 &   327.011 &  $-$7.556 & 5BZB J2148$-$0733 &   1.2 &  6 & 4.508 & 27.047 & 0.029  &  D3PO$^4$ \\ 
MST 2156$-$0037 &   329.087 &  $-$0.625 & 5BZU J2156$-$0037 &   1.3 &  9 & 4.122 & 37.096 & 0.078  &  \\
MST 2254$-$2725 &   343.733 & $-$27.415 & 5BZB J2254$-$2725 &   0.6 &  7 & 4.112 & 28.783 & 0.049  &  \\
\hline
MST 0022$+$0008 &     5.492 &   0.135   & 5BZG J0022$+$0006 &   1.3 & 10 & 2.629 & 26.291 & 0.136  & 2FHL$^a$ \\  
MST 0115$-$3400 &    18.742 & $-$34.004 & 5BZB J0115$-$3400 &   0.8 &  9 & 4.786 & 43.077 & 0.032  & 2FHL$^b$ \\  
MST 0304$-$0055 &    46.160 &  $-$0.921 & 5BZB J0304$-$0054 &   1.6 &  8 & 3.515 & 28.122 & 0.094  & 2FHL$^c$ \\  
\hline
\end{tabular}
\begin{flushleft}
{\footnotesize 
$^\dagger$1FGL J0722.3+5837. See text for discussion. \\
$^1$1DF163831; $^2$1DF145964; $^3$1DF022985; $^4$1DF164164 \\
$^a$2FHL~J0022.0$+$0006; $^b$2FHL~J0114.9-3359; $^c$2FHL~J0304.5-0054; 
}
\end{flushleft}

}
\label{tab:testfieldsrc}
\end{table*}

\section{The MST cluster populations}\label{s:clusters}

As reported in Paper II, \emph{Fermi}-LAT data (Pass 8) above 10~GeV, covering the 
whole sky in the 7 years time range from the start of mission (2008 August 04) 
up to 2015 August 04, were downloaded from the FSSC 
archive\footnote{\url{http://fermi.gsfc.nasa.gov/ssc/data/access/}}.
Standard cuts on the zenith angle (100\degr) and data quality were applied.

The search for clusters of $\gamma$-ray photons by means of MST was performed after the 
exclusion of the Galactic belt up to a latitude $|b| \leq 25^{\circ}$ to reduce the
possibility of finding clusters originated by local high background fluctuations.
Each of these two spherical broad regions was then divided into ten smaller parts
where MST was applied.
The parameters of primary selection of clusters in the 10 GeV sky were $N_\mathrm{cut} = 4$
and $\Lambda_\mathrm{cut} = 0.7 \Lambda_{m}$; then a secondary selection was applied with 
a rather robust threshold, $M > 20$. 
A sample of 921 clusters was obtained, of which 716 have a firm 2FGL or 3FGL counterpart and one is coincident
with the well known GRB 130427A \citep{maselli14,ackermann14}. For
165 clusters 1FHL counterparts were found, five of which not included in the 3FGL catalogue.
Therefore, the remaining sample contains 199 clusters, of which 189 are not related to previously known $\gamma$-ray
sources and 10 are associated with the very recently published 2FHL catalogue \citep{2FHL} at energies higher than 50 GeV, 
also based on the Pass 8 sky.

\begin{table*}[htb!]
\caption{Standard unbinned likelihood analysis of the \emph{Fermi}-LAT data, see 
Sect. 4 for details. 
The third and fourth columns report photon fluxes in units of 
10$^{-11}$ ph\,cm$^{-2}$\,s$^{-1}$.  
For sources below the usual significance threshold ($\sqrt{TS}=5$) only upper 
limits are given.
Radio, optical data and redshift from 5BZCAT are also given.}
\label{table2}
\centering
{
\begin{tabular}{ccrccrrr}
\tableline
 MST cluster &  $\sqrt{TS}$  &    Flux~~    &    Flux     & Photon index  &  $F_{1.4}$ &  $R$  &  $z$~   \\
        &               & 3--300 GeV   & 10--300 GeV &               &    mJy     &       &         \\
\tableline                                                 
MST 0216$+$2314 & 6.4 & $10.0\pm2.8 $ & $3.3\pm1.1$ & $1.9\pm0.3 $ &  36 & 18.4 & 0.288 \\  
MST 0219$-$1725 & 8.3 & $10.0\pm2.4 $ & $2.9\pm1.0$ & $2.0\pm0.3 $ &  62 & 17.1 & 0.128 \\  
MST 0314$+$0619 & 9.2 & $13.0\pm2.9 $ & $3.7\pm1.2$ & $2.0\pm0.3 $ &  29 & 17.5 &   --- \\  
MST 0503$-$1115 & 8.0 & $10.0\pm2.5 $ & $3.3\pm1.1$ & $1.9\pm0.3 $ &  10 & 19.9 &   --- \\  
MST 0712$+$5719 & 4.3 & $< 4.9$      & $< 1.0$     & ---           &   8 & 20.1 & 0.095 \\  
MST 0723$+$5841 & 4.6 & $< 4.7$      & $< 1.0$     & ---           &  16 & 18.2 &   --- \\  
MST 0820$+$2353 & 6.3 & $5.0\pm1.7$  & $2.0\pm0.8$ & $1.7\pm0.3 $  &  49 & 18.0 & 0.402 \\  
MST 0828$+$2312 & 4.2 & $< 4.3$      & $< 1.4$     & ---           &  36 & 19.9 &   --- \\  
MST 0913$+$8133 & 4.8 & $< 5.2$      & $< 1.1$     & ---           &   5 & 20.4 & 0.639 \\  
MST 1055$-$0126 & 8.5 & 9.0$\pm$2.5  & 4.2$\pm$1.3 & 1.6$\pm$0.2  &  12 & 18.5 &   --- \\   
MST 1105$+$3946 & 6.1 & 5.8$\pm$1.9  & 1.7$\pm$0.8 & 2.0$\pm$0.4  &  43 & 15.1 & 0.099 \\   
MST 1134$-$1729 & 4.8 & $< 5.5$      & $< 1.7$     & ---           &   5 & 19.2 & 0.571 \\  
MST 1208$+$3015 & 6.1 & 6.1$\pm$1.8  & 2.2$\pm$0.8 & 1.8$\pm$0.1  &  36 & 22.1 &   --- \\   
MST 1215$+$0732 & 6.7 & 6.8$\pm$2.1  & 2.3$\pm$0.9 & 1.9$\pm$0.3  & 137 & 16.4 & 0.136 \\   
MST 1402$+$1559 & 4.5 & $< 3.1$      & $< 0.9$     & ---           & 849 & 17.7 & 0.244 \\  
MST 1449$+$2746 & 6.3 & 4.1$\pm$1.5  & 2.1$\pm$0.9 & 1.5$\pm$0.4  &  91 & 18.2 & 0.227 \\ 
MST 1506$-$0540 & 6.1 & 6.7$\pm$2.3  & 1.8$\pm$0.8 & 2.1$\pm$0.4  &  15 & 19.1 & 0.518 \\   
MST 1515$+$2426 & 5.1 & 3.8$\pm$1.5  & 1.7$\pm$0.8 & 1.6$\pm$0.4  &  34 & 17.9 & 0.228 \\
MST 1518$+$4045 & 7.7 & 8.8$\pm$2.2  & 2.0$\pm$0.8 & 2.2$\pm$0.3  &  44 & 15.1 & 0.065 \\ 
MST 2148$-$0733 & 7.6 & 8.9$\pm$2.4  & 2.6$\pm$1.0 & 2.0$\pm$0.3  &   8 & 18.4 &   --- \\   
MST 2156$-$0037 & 9.3 & 17.0$\pm$3.3 & 3.4$\pm$1.1 & 2.3$\pm$0.3  & 233 & 20.9 & 0.495? \\
MST 2254$-$2725 & 9.6 & 14.0$\pm$3.0 & 3.3$\pm$1.1 & 2.0$\pm$0.3  &  53 & 17.8 & 0.333 \\   
\hline
MST 0022$+$0008 & 6.9 & $6.0\pm3.1  $ & $3.2\pm1.5$ & $1.4\pm0.7 $ &   2 & 18.6 & 0.306 \\  
MST 0115$-$3400 & 8.4 & $5.8\pm1.7  $ & $3.2\pm1.1$ & $1.4\pm0.1 $ &   6 & 19.1 & 0.482 \\  
MST 0304$-$0055 & 6.3 & $6.7\pm2.2  $ & $2.5\pm1.0$ & $1.8\pm0.3 $ &  28 & 18.6 & 0.511 \\  
\hline
\end{tabular}
}
\end{table*}

The superposition of the positions of these 199 clusters with the 5BZCAT blazars
with a positional matching within a maximum angular separation, computed using the cluster 
centroids' coordinates, of 6\arcmin\ (0\fdg1, this value was found in Paper I to be optimal),
provided 34 clusters: 9 of them were already found in Paper I using the
Pass 7 sky\footnote{In that paper a more loose selection threshold was used, $M>15$. See also Section \ref{s:discussion} for a discussion.} 
and the remaining 25 new clusters are listed with their most important parameters 
in Table 1.  

Three of these 25 clusters were found to match new sources in the 2FHL catalogue \citep{2FHL} at energies higher than 50 GeV.
Their angular distance is lower than 2\arcmin\ and therefore these associations can be considered    
as safe.
These 3 sources are reported in the lower section of Table 1, where their 
2FHL counterpart is also shown.
Therefore, a total of 22 clusters associated with blazars not previously known as $\gamma$-ray emitters is found.
It should be pointed out that, as shown in Table \ref{table1}, 
4 of these clusters are in 
good positional coincidence (between 0\fdg05 and 0\fdg24) with sources included in the D3PO\footnote{The Denoised, Deconvolved, 
and Decomposed Fermi $\gamma$-ray Sky, \url{http://www.mpa-garching.mpg.de/ift/fermi/}. 
Note that this catalogue uses data starting from 600 MeV, where the PSF and the source localisation is worse.} catalogue \citep{selig14}.

In the 5BZCAT the large majority (18) of these associated blazars are classified 
as BL Lac objects (BZB), six are classified as galaxy-dominated blazars (BZG) while one is 
a blazar of uncertain classification (BZU).

Remarkably, no Flat Spectrum Radio Quasar (BZQ), the most numerous class of
blazars in the 5BZCAT with 1909 objects over a total of 3561 sources ($\sim$54\%), was 
found.
If the MST-5BZCAT associations were due to chance, it could be expected that more than 50\% of blazars
should be BZQ.
A simple calculation of the binomial probability to have zero positive cases over 25 with
a probability of 1909/3561 = 0.536 gives the very low value of
$\approx 5 \times 10^{-9}$.

An estimation of the expected number of correspondences $N_\mathrm{ex}$ assuming a random matching
between MST cluster centroids and catalogued sources can be 
derived (see also Paper II) from the ratio between the sum of solid angles of clusters $\omega_i$ 
to the total solid angle of the explored sky region $\Omega$, multiplied by 
the number of blazars $N_b$:
\begin{equation}
N_\mathrm{ex} = N_b \frac{\sum_i \omega_i}{\Omega} = N_c N_b \frac{\omega}{\Omega}
\end{equation}
where $N_c$ is the number of clusters, assumed all having the same solid angle $\omega$
corresponding to a circular region of radius equal to 0\fdg1.
From $\Omega = 7.256$ sr and $\omega = 9.57 \times 10 ^{-6}$ sr, with $N_c = 199$
and $N_b = 1652$, that is the number of objects in the 5BZCAT excluding FSRQs (BZQ),
we obtain $N_\mathrm{ex} = 0.43$, a value much smaller than our result.

\section{Maximum likelihood analysis}\label{s:maxlike}

Similarly to the analysis performed in Papers I and II, a standard unbinned likelihood 
analysis was done for each MST cluster. 
The Region of Interest (ROI), with a radius of 10\degr, is centered at the MST 
cluster centroid coordinates, and standard screening criteria were applied to the 
\emph{Fermi}-LAT data above 3 GeV. 
This lower energy threshold was chosen in order to confirm the detection of sources 
also at lower energies, where MST is less efficient for selecting clusters because of the much higher 
photon density and the presence of a rich population of bright sources.
The likelihood analysis was performed considering all the 3FGL sources within 
20\degr\ from the cluster centroid, as well as the Galactic and extragalactic 
diffuse emission. 
A further source with a power-law spectral distribution was assumed at the MST 
coordinates. 
The normalization and spectral index of all the 3FGL sources within the ROI 
was allowed to vary in the fitting, while the parameters of the sources between 
10\degr\ and 20\degr\ from the center of the field of view were fixed to their 
catalogue values.
The likelihood Test Statistics ($TS$) was derived from this analysis, together with 
the fluxes in the two 3--300 and 10--300 GeV bands for the sources with $\sqrt{TS} > 5$.
The results are reported in Table~\ref{table2}. 

\begin{figure*}
\centering
\includegraphics[width=0.9\textwidth]{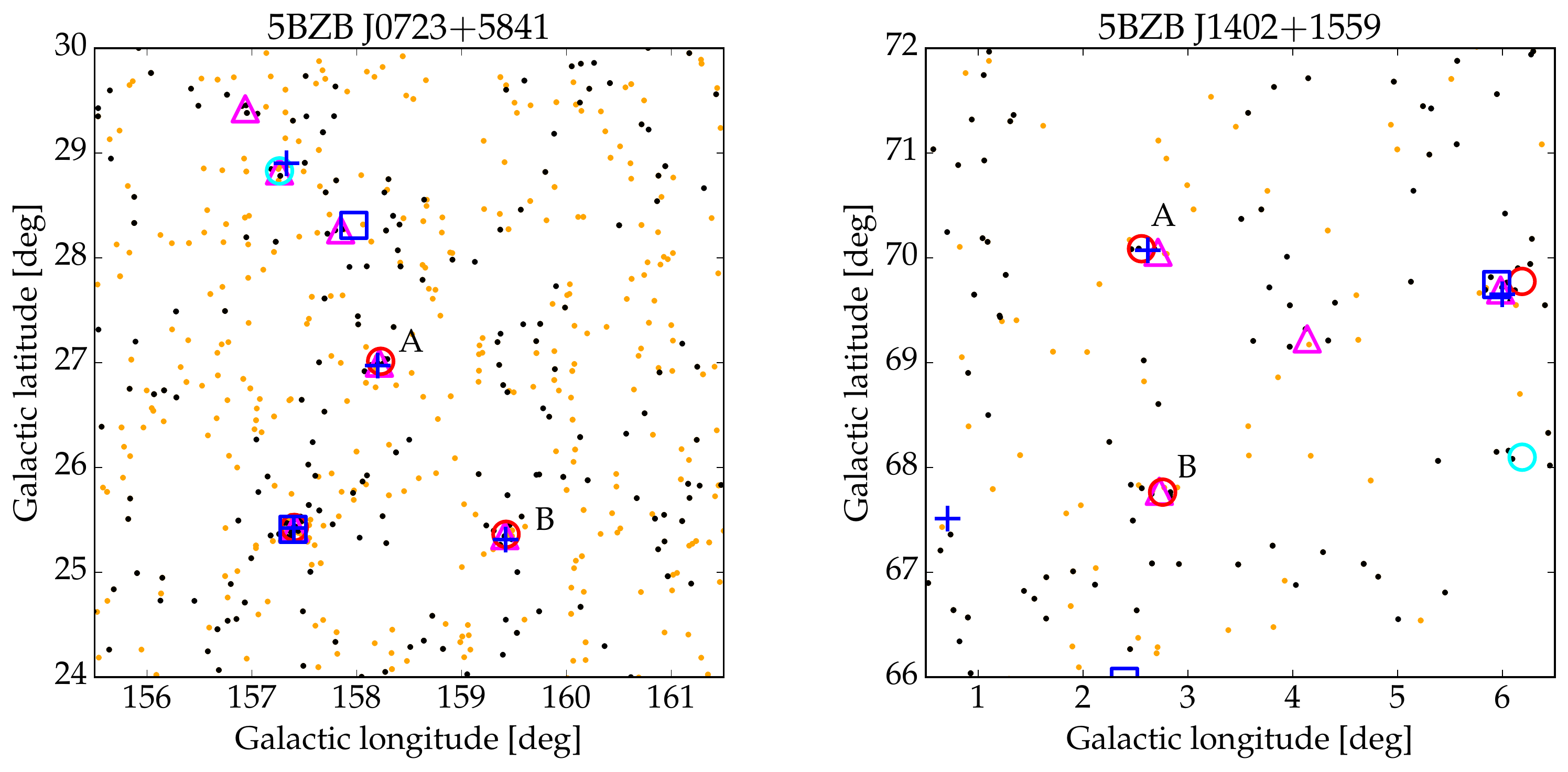}
\caption{
Photon sky maps in Galactic coordinates of regions containing the BL Lac objects 
5BZB J0723$+$5841 (left panel), and 
5BZB J1402+1559 (right panel).
Black filled circles mark the photon coordinates at energies higher than 10 GeV, 
\textbf{orange} points are photons with energies in the range [7, 10] keV;
large black crosses correspond to the optical coordinates of BL Lac objects in the 5BZCAT 
blue open squares are the 3FGL sources, 
open red  circles are the MST ($>$10 GeV, $M > 20$) cluster positions, 
\textbf{cyan} open circles are lower-significance clusters with $12 < M < 20$ 
and triangles are clusters detected at energies higher than 7 GeV ($M > 15$).
In each panel the letter A marks the 5BZCAT source. 
The letter B in the left panel indicates the cluster corresponding to 5BZB J0712$+$5719 and in the right panel another interesting cluster 
discussed in the text and up to now not associated with a known blazar or $\gamma$-ray source.
}
\label{BZB_tre}
\end{figure*}

ML analysis thus confirmed all these findings and only six objects have 
$ 4 < \sqrt{TS} < 5$: two of them have a high clustering with $g > 3$,
whereas for the other four clusters it is lower.
In the following subsections we discuss about the actual significance of these detections.

Finally, the aperture photometry lightcurve was computed for each cluster. 
No significant variability was detected for any source. 

\subsection{High $g$ clusters with low $TS$}

The two panels in Figure~\ref{BZB_tre} show the 6\degr$\times$6\degr\ regions containing the
two low-significance clusters that have high $g$ values (MST~0723$+$5841 and MST~1402$+$1559): 
in both panels all the known BL Lac objects correspond to photon clusters detected at energies 
above 10 GeV and also above 7 GeV.

In the left panel three blazars corresponding to clusters are visible: 
5BZB J0723$+$5841 is located close to the most significant cluster (marked with ``A'' in the figure);
it was already reported in the 1FGL catalogue with a statistical significance of 4.56$\sigma$ and a photon flux of
$(6.26 \pm 2.66) \cdot 10^{-10}$ ph~cm$^{-2}$~s$^{-1}$ but not confirmed in the 2FGL and 3FGL catalogues.
The other two blazars were previously undetected in the $\gamma$-ray band:
5BZB J0712$+$5719 is very close to the low $g$ cluster MST 0712$+$5719 (marked with ``B'', see next Section)
and 5BZG J0737$+$5941 is near a cluster detected above 7 GeV but below our $M$ threshold 
at higher energy.
The latter object is a flat spectrum radio galaxy (UGC 3927, S4 0733$+$59) at the 
redshift $z = 0.0405$ already considered as a possible GeV-TeV source
\citep{abdo09}.

The region on the right panel contains three known BL Lacs of which only one was
already reported as a 3FGL source.
One of the other two corresponds to the cluster found in this work, while no
apparent concentration of high energy photons is close to the other 5BZCAT object.

It is interesting to note that, in addition to two low significance clusters,
there is another one with 7 photons ($>$10 GeV) but a very high $g = 5.259$ and
$M = 36.811$ (marked as ``B'' in the map) without an obvious counterpart.
The centroid of this cluster is at only $\sim$20\arcsec\ from the flat spectrum radio
source TXS 1408$+$148, with the optical counterpart SDSS J141028.05+143840.1 a galaxy
at $z = 0.144$ and proposed as a $\gamma$-ray candidate by \cite{dabrusco14}
on the basis of its WISE colours.
We found several other clusters with similar characteristics and not associated
to known blazars or candidate, and this topic will be discussed in a forthcoming
paper.

It appears, therefore, that the associations of these three low-$TS$, high-$g$ clusters with BL Lac
objects could be considered safe. In particular, the early 1FGL $\gamma$-ray detection of 
5BZB J0723$+$5841 is now confirmed by the well apparent photon cluster above 10 GeV in Figure~\ref{BZB_tre}.

\subsection{Low $g$ clusters with low $TS$}

Clusters with $g$ values lower than 3 can reach a significance above the $TS$ threshold,
as in the cases of MST 1449$+$2746 (associated with 5BZG J1449$+$2746) and MST 0022$+$0008 (5BZG J0022$+$0006 and 2FHL J0022.0+0006), 
but more frequently they are also found with $\sqrt{TS} < 5$.
This fact can be interpreted in different ways, considering that their photon number must
generally be higher than 8 to reach the $M$ threshold for the secondary selection.
The low clustering can be thus due either to a small dense core in a low density halo
or to a rather sparse cluster structure.
In any case, these clusters do not have a good \emph{stability}, in the sense that a small
change of the selection parameters, as the size of the search region (affecting the
value of $\Lambda_m$) or the adopted $\Lambda_\mathrm{cut}$, can produce a variation of $M$ across
the threshold.
These clusters might be related to weak sources but the possibility of
a local background fluctuation cannot be excluded.    
In any case, an inspection with a finer study of the their environment could help to
understand their nature.

The cluster MST 0712$+$5719 corresponding to the BL Lac object 5BZB J0712$+$5719, that has the lowest $g$, 
is indicated by the letter ``B'' in the photon map in the left panel of Figure~\ref{BZB_tre}.
As noticed above all the clusters at energies higher than 10 GeV present in this region
were found very close to blazars and thus it appears very unlikely that random background
fluctuations have such a good positional matching.
In these 6\degr$\times$6\degr\ fields the only clusters present are all coincident with
BL Lac objects.

In the field of MST 0828$+$2312, whose $g$ is close to 3, there are only three 
clusters above the $M$ threshold (one corresponding to a 3FGL source and the other to
5BZB J0820$+$2353) and all are matching BL Lac positions.
There is only one cluster (MST 0913$+$8133, $M > 20$) in a 6\degr$\times$6\degr\ field around 5BZB J0913$+$8133 with a very
good positional correspondence.
Again, three clusters ($M > 20$) are in the last 6\degr$\times$6\degr\ field: one is MST~1134$-$1729 
associated with 5BZB J1134$-$1729, another matches a 3FGL source 
and the other corresponds to 5BZB J1137$-$1710. All these clusters in a very good positional agreement
with BL Lac objects.

In conclusion, the analysis in small sky regions around all these objects demonstrated that
all the selected clusters are related to astrophysical objects candidate to be $\gamma$-ray
sources and the possibility that they are due to background density fluctuations at a so high
energy could be considered negligible for all practical purposes.

\section{Properties of the newly detected blazars}\label{s:properties}

In Table \ref{table2} we reported in addition to $\gamma$-ray significance and spectral data
some other interesting parameters: the radio flux density at 1.4 GHz
from NVSS or FIRST, optical magnitude in the $R$ or $r$ (SDSS) bands and the redshift 
$z$, useful for the understanding of the main properties of these blazars.

Two of the BL Lac objects found in this search are classified as candidate in the 5BZCAT,
while the other 16 are confirmed.
Optical spectra of the latter type are therefore available in the literature or in the
web, while those of the former ones are unpublished.
Spectra are generally dominated by a continuum with a blue excess typical of HBL objects.
14 of the 18 BL Lacs are also in the 1WHSP catalogue \citep{arsioli15} supporting this classification.  

\begin{figure}[h!]
\centering 
\includegraphics[width=0.47\textwidth]{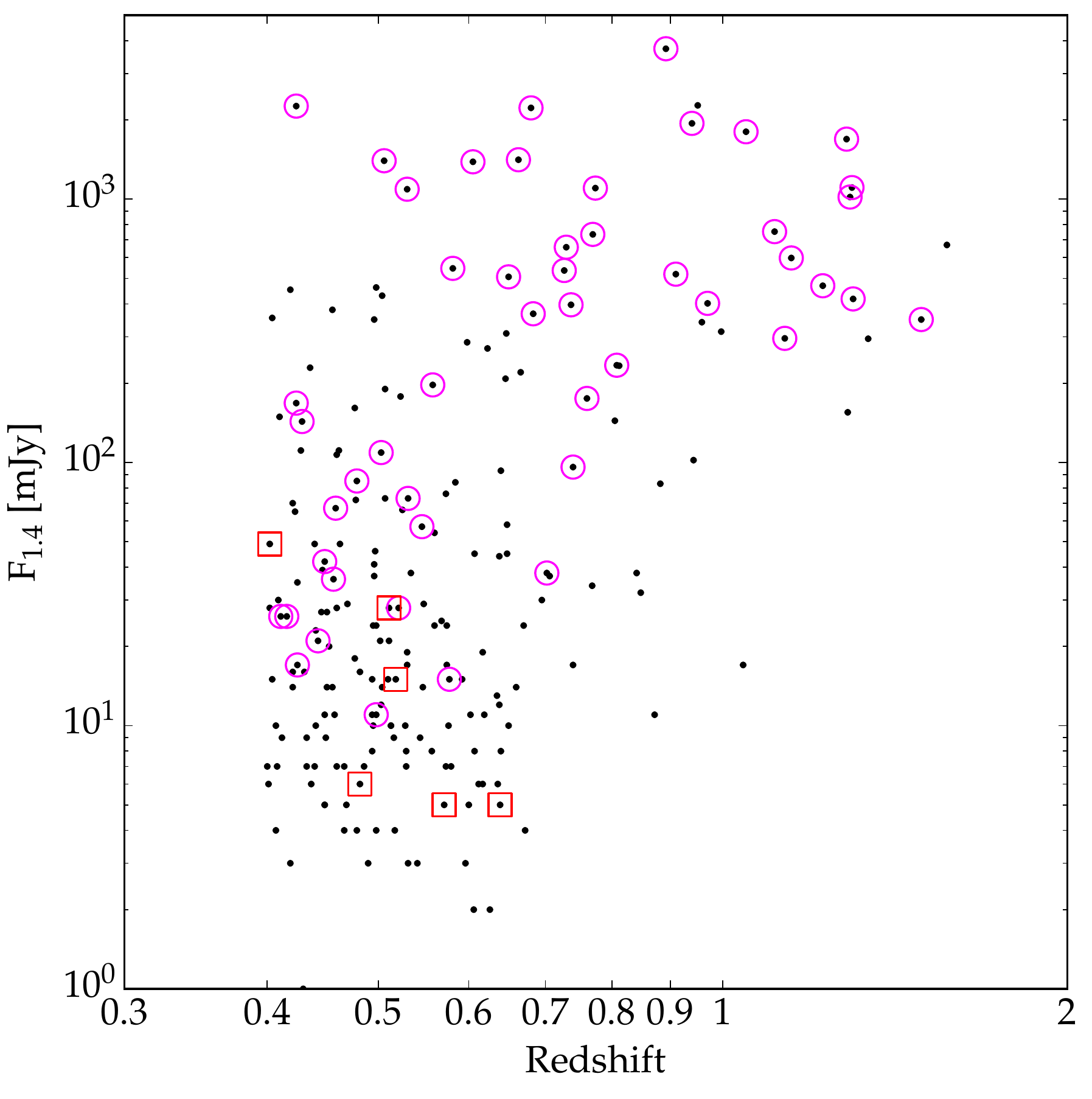}
\caption{Radio flux density at 1.4~GHz vs redshift for the BZB sources with $z > 0.4$
in the 5BZCAT (black points);
large magenta circles mark sources previously detected at $\gamma$-ray energies
while red squares mark the seven sources found in the present work.
}
\label{hzBL_radio}
\end{figure}

Note that there are six BL Lac objects at $z > 0.4$, two of which are also 
in the 2FHL catalogue.
In the 5BZCAT redshifts are reported for 437 BL Lac objects and 208 have a value higher 
than 0.4.
Figure \ref{hzBL_radio} shows the plot of the radio flux density at 1.4 GHz as a function of $z$ for 
the subsample of the latter sources.
At radio flux densities higher than 100 mJy a large fraction of BL Lac objects was 
already detected in the $\gamma$-rays (magenta circles) whereas below this value 
the number of detections decreases rapidly and none is present below 10 mJy. 
Three of our six sources are at flux levels under this limit.
This result indicates that our findings increased the number of low luminosity sources
detected at these energies.

In the following subsections the properties of some individual sources, in particular those
classified as BZG and BZU, are discussed.

\subsection{5BZB J1402$+$1559}
 
This source has the highest flux density in our sample and its optical spectrum 
presents either emission lines with a borderline equivalent width $EW \approx 6$ \AA\ 
between the typical values for BL~Lacs and quasars or a featureless power law continuum 
\citep{falomo94}.
In the 5BZCAT it is listed among BZB sources but a note claims for a possible 
BZU classification.
A VLA image at 20 cm \citep{antonucci85} reveals an extended complex 
structure with a dominant compact core and the radio spectrum can result steep 
according to the used resolution.
In \cite{dabrusco14} it was proposed as a candidate $\gamma$-ray source because 
its WISE mid-infrared colours are similar to those of confirmed $\gamma$-ray 
blazars.
The present detection confirms these expectations for the occurrence of a blazar 
core.

\subsection{5BZU J2156$-$0037}
This optically faint source was classified of {\it uncertain} type in the 5BZCAT 
because the available spectra \citep[][SDSS]{hook03} are noisy and the occurrence 
of weak emission lines could not be excluded.
The redshift $z = 0.495$ given by \cite{jackson02} is therefore unsafe.
It is possible that this source is a far BL Lac object similar to the other ones
found in this search, and new spectroscopic analysis will be useful to establish
its nature.
In any case the detection in the high energy $\gamma$-ray band is strongly supported
by the very significant $\sqrt{TS} = 9.3$.

\subsection{The 5BZG sources}
The 5BZCAT reports 274 BZG sources and only 27 of them ($\sim$10\%) are associated with $\gamma$-ray
sources.
In Table \ref{table1} there are 6 objects of this type with redshifts in the range from 0.065 to
0.306; all were confirmed with $\sqrt{TS} > 5.0$ by the maximum likelihood analysis.
Some of them exhibit characteristics closer to radio galaxies than BL Lacs and therefore
are presented in some detail in the following.

\subsubsection{5BZG J0022$+$0006}

In the present sample this object is the faintest radio source.
It was first indicated as a possible BL Lac counterpart to a ROSAT source by 
\cite{brinkmann00} and later by \cite{collinge05} using SDSS spectra.
The WISE colours ($W1 - W2 = 0.34, W2 - W3 = 1.97$) are compatible with the blazar 
region defined by \cite{dabrusco14}.
Its spectrum in SDSS, however, presents a clear dominance of the host
galaxy emission with a well evident Ca H\&K break.
There is also a very close $\gamma$-ray source in the 2FHL catalogue at energies above 50~GeV \citep{2FHL}. 

\subsubsection{5BZG J1105$+$3946}

\begin{figure}[h!]
\centering 
\includegraphics[width=0.48\textwidth]{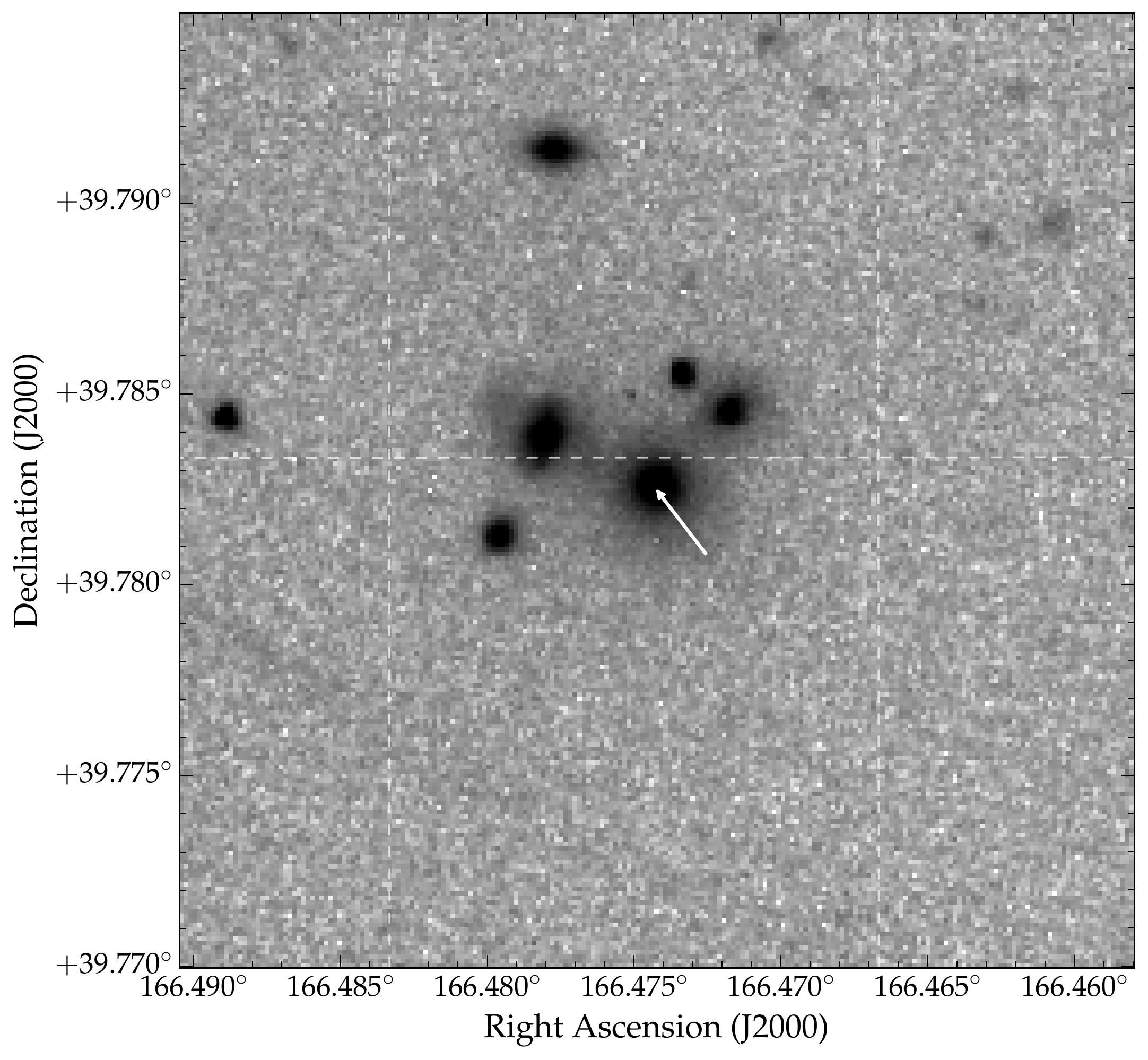}
\caption{Image of the compact group of galaxies Shk~007 with 5BZG~J1105$+$3946 
indicated by the arrow (adapted from SDSS $r$-band image). The image side is 1\farcm5.}
\label{shk007}
\end{figure}

This source is in the compact group of galaxies Shk~007 
\citep[][see also Figure~\ref{shk007}]{shakhbazyan73,stoll96}.
It is a CLASS source reported as a blazar by \cite{marcha01} and is also in
the Low-frequency Radio catalogue of Flat-spectrum Sources by \cite{massaro14a}.
The SDSS optical spectrum is typical of an elliptical galaxy with some narrow 
emission lines.
No X-ray detection is reported.
WISE colours ($W1 - W2 = 0.41$, $W2 - W3 = 2.27$) are marginally compatible with the 
\emph{locus} occupied by $\gamma$-ray blazars in the colour space \citep{dabrusco14}.
There is a Planck source at about 2\farcm2 \citep{planck14}.
This source appears likely as an active radio galaxy rather than a typical BL Lac 
object.

\subsubsection{5BZG J1215$+$0732}
This source, also known as 1ES 1212+078, is apparently located inside the galaxy 
cluster GMBCG J183.79574+07.53462, whose spectroscopic redshift is not known but 
the photometric estimate is 0.227, that if confirmed is about 70\% higher than the one of the 5BZG source.
It is generally reported in the literature as a BL Lac object, but the optical 
emission in the SDSS spectrum appears dominated by the host galaxy component.
A VLA radio image \citep{rector03} shows a jet elongating for about 50 pc 
and in \cite{massaro13} it was proposed as a TeV emitter candidate on the basis of 
ROSAT and WISE data.
The present detection above 10 GeV confirms the BL Lac properties of this source.

\subsubsection{5BZG J1449$+$2746}

This BZG object is the closest to the MST cluster centroid position.
It is a good candidate as high energy source because it was identified as the 
counterpart to a ROSAT source and is classified as a HBL.
However, this is a case of possible confusion because there is another object
that could be also associated with the $\gamma$-ray emission.
The closer and brighter galaxy MCG $+$05-35-018 ($z = 0.030$) is at only 
1\farcm20; it is also reported in the CRATES catalogue because of its flat radio 
spectrum and is considered variable by \cite{thyagarajan11}.
Moreover, there is the galaxy cluster MSPM 00116 ($z = 0.031$), that could be
partially responsible for X and radio emission.

\subsubsection{5BZG J1515$+$2426}

This radio source was classified as a BL Lac object by \cite{brinkmann00}, but in SDSS it 
appears a red galaxy ($u - r = 1.88$).
No other interesting possible counterpart to the $\gamma$-ray cluster can be found up to 
a 12\arcmin\ radius.

\subsubsection{5BZG J1518$+$4045}

This flat spectrum radio source was identified as the counterpart of a ROSAT source 
\citep{laurent97,brinkmann00} and reported as a weak line AGN 
in the CLASS BL Lac sample \citep{marcha13}. 
The SDSS spectrum of this galaxy presents several emission lines (H$\alpha$, N\textsc{ii}, S\textsc{ii}, 
O\textsc{ii}), the high Ca H\&K break and the colour $u - r = 2.60$ that justify its classification 
as a BZG source in the 5BZCAT, while in the previous editions it was named as BZU.
Moreover, the WISE colours are largely outside the \emph{locus} occupied by $\gamma$-ray 
blazars in the colour space \citep{dabrusco14} and close to the regions of stars 
and normal galaxies.
This source, therefore, could be a transition object between a radio galaxy 
and a blazar.

\section{Summary and discussion}\label{s:discussion}

Our application of MST for searching localized photon clusters in the LAT Pass 8
sky confirmed the good performances of this method.
We verified also the consistency of the Paper I (Pass 7) sources with the Pass 8 sky:
15 of the 19 detected clusters in the older data were confirmed with a comparable or higher 
significance; in some cases a reduced number of photons was overcompensated by a
higher clustering parameter. Two the remaining 4 clusters (MST~0932+1042 and MST~1005+6443) were 
found with a low photon number but with $g > 3$, while the significance of other two 
(MST~1311+3951 and MST~1423+1414) decreased because both $n$ and $g$ resulted lower 
than in Pass 7. 
To gain more confidence on the cluster significance in the present work we applied 
a secondary selection using higher values of the parameters, as already done in Paper II.

In the present paper we reported 25 new clusters to be associated with known 
BL Lac or similar objects (3 of them are in the new 2FHL catalogue). 
No association with FSRQ sources was found, supporting the picture of a GeV to 
TeV sky preferentially dominated by BL Lacs.
This result can be understood  on the basis that BL Lac objects, and particularly HBLs, 
have gamma ray spectra much harder than FSRQs, as clearly apparent in the photon index
plot by \cite{singal12}.
In the 3LAC population \citep{ackermann15}
the mean photon index of BL Lacs is $2.01\pm0.25$, and the one of HBLs is $1.87\pm0.20$, 
while FSRQs have $2.44\pm0.20$. 
Therefore it can reasonably be expected that above 10 GeV the new and weaker sources, not already detected 
at lower energies, are mostly HBL objects, and this is also supported by the values
of the photon indices in Table \ref{table2}.

Our detections were generally confirmed by the standard maximum likelihood analysis.
Considering also the 19 BL Lacs previously discovered in the Pass 7 sky (Paper I)
and the other 16 sources (Paper II) in the 1WHSP catalogue \citep{arsioli15}, 
the use of our method based on the MST clustering has provided up to now several tens of new possible
$\gamma$-ray blazars.
It should be emphasized, however, that these discoveries are mainly due to the improvement of
instrumental response functions used for producing the Pass 8 sky and to an exposure with a duration
of about twice the one considered at the epoch of 3FGL catalogue.

\begin{acknowledgements}
We acknowledge use of archival Fermi data. We made large use of the online version of the Roma-BZCAT 
and of the scientific tools developed at the ASI Science Data Center (ASDC),
 of the Sloan Digital Sky Survey (SDSS) archive, of the NED database and other astronomical 
catalogues distributed in digital form (Vizier and Simbad) at Centre de Dates astronomiques de 
Strasbourg (CDS) at the Louis Pasteur University.
\end{acknowledgements}

\bibliographystyle{spr-mp-nameyear-cnd}
\bibliography{bibliography} 

\end{document}